# Re-examining the Philosophical Underpinnings of the Melting Pot vs. Multiculturalism in the Current Immigration Debate in the United States

**Daniel Woldeab**
College of Individualized Studies, Metropolitan State University,
St. Paul, MN, USA
Daniel.Woldeab@metrostate.edu
https://orcid.org/0000-0002-8267-7570

**Robert M. Yawson**
School of Business, Quinnipiac University
Hamden, CT, USA
robert.yawson@qu.edu
https://orcid.org/0000-0001-6215-4345

**Irina M. Woldeab**
Minnesota Department of Natural Resources,
St. Paul, MN, USA
imwoldeab@gmail.com

# Abstract

Immigration to the United States is certainly not a new phenomenon, and it is therefore natural for immigration, culture and identity to be given due attention by the public and policy makers. However, current discussion of immigration, legal and illegal, and the philosophical underpinnings is 'lost in translation', not necessarily on ideological lines, but on political orientation. In this paper we reexamine the philosophical underpinnings of the melting pot versus multiculturalism as antecedents and precedents of current immigration debate and how the core issues are lost in translation. We take a brief look at immigrants and the economy to situate the current immigration debate. We then discuss the two philosophical approaches to immigration and how the understanding of the philosophical foundations can help streamline the current immigration debate.

# Immigration to the United States

Immigrants are certainly not a new phenomenon to the United States of America, and it is, therefore, natural for immigration, culture and identity to be given due attention by the public and policy makers. In fact, "The United States of America has been the preferred destination for immigration since the discovery of the New World" (McGruder, 2016, p. 1). As such, feelings around immigrants and approaches to immigration policy have gone through many phases and iterations (Woldeab, Yawson, & Woldeab, 2019).

It seems, however, that "once again, the United States finds itself in an era of nativism and exclusion, as our politicians contemplate immigration restrictions and deportation policies that are reminiscent of those enacted nearly a century ago" (Young, 2017, p. 218). Indeed, over the last few years the media has sensationalized the irregular migration from what is considered the 'third world' to well-to-do western nations (Flahaux & Haas, 2016), while lawmakers and politicians alike "frequently portrayed immigration as a threat to the nation" (Young, 2017, p 218). This is not new. Indeed, the country has a fraught relationship with its own views about the immigrants who founded the United States in ways that dominated over, excluded, and actively marginalized Native American populations, and the varying views of later inflows of immigrants. For over a century, however, two opposing schools of thought took shape and have driven the process of acculturation (multiculturalism). In light of the current policy environment, framed by the Trump administration's motto "Make America Great Again", this paper discusses the philosophical underpinnings, as well as some of the outcomes, of those two schools of thought.

At the turn of the century, Rudmin (2003) noted that the issue of acculturation is becoming increasingly important for converging reasons: high-speed, high-volume new technologies; a flow of millions of new immigrants as a result of war, political oppression, economic disparities, and environmental pressures; and regional and global free-trade arrangements that resulted in international marketing and recruitments for skilled personnel. Chiswick and Hatton (2003) further stated that the characteristics of the late 20th century would continue into the 21st century with an increase in the movement of people across international borders. For Nee and Sanders (2001) the immigration of people to advanced capitalist societies is among the major societal movements of our time. Furthermore, Rosenblum (2017) noted that today's immigration flows are much more complex, not just because of technology and lower travel costs, but also because they come on the foundation of previous immigration flows.

As far as the size of immigration flows goes, Martin (2016) notes that while the United Nations report counts 244 million immigrants worldwide, "The US is the nation of immigration, with almost 20 percent of the world's international migrants and half of the unauthorized migrants in industrial countries" (p. 295). According to Perlstein (2017), the United States welcomed 59 million immigrants between 1965 and 2015. In fact, the National Academies of Sciences, Engineering, and Medicine (2007) asserted that today more than 40 million individuals who call the United States home were born in other countries, and roughly an equal number of people have at least one foreign-born parent; this means one person out of five is either first generation foreign born or their children. It is thus not surprising that immigration is front and center on the national policy agenda.

In terms of the composition of those immigration flows, while in 1980 there were only 816,000 immigrants who identified themselves as black in the United States, that number quickly climbed to 4.2 million by 2016, with most of the growth taking place since 2000 (Anderson & López, 2016). Likewise, Antonio Flores (2016) of Pew Research Center reported that the Latino population in the US was approaching 58 million by 2016. There are also about 11 million undocumented residents living with almost 6 million US-born children (Warren & Kerwin, 2017). Overall, Martin (2016) reported that the majority of immigrants in the United States are from Latin America and a fourth are from Asia. What seems to matter most however – and perhaps this has always been the case – is the perceived integration of current immigrants.

Indeed, "for neo-conservative wings of the New Right, the questions of how many or what color are less important than the issue of the potential for recent immigrants to assimilate into the mainstream economy and society, as did previous waves of immigrants" (Ansell, 2016, p. 126-127). Although each immigration wave tends to trigger worries about integration, national identity, cultural integrity, crime, and job stability among others, the period we are in now started long before the November 2016 election and is marked by seemingly sharper differences between advocates of immigration restrictions and those who hold more liberal views on the topic (Martin, 2016). Indeed, as "a nation of immigrants" the United States remains "unsure about the best migration policy for the future" (Martin 2016, p. 295). This is particularly clear in the cyclical views of the relationship between immigration and the economy, which is the subject of the next section.

## Immigrants and the Economy

The United States – which was built on the backs of first people, slaves, and immigrants – has long seen immigrants both as an opportunity to develop and advance its economy and also as a burden on its social services. Indeed, Abramitzky and Boustan (2016) asserted that "the history of immigration to the United States has been shaped both by changes in the underlying costs and benefits of migration as well as by substantial shifts in immigration policy" (p.5).

McGruder (2016) noted that opinions are very much divided: some see immigrants as taking a toll on social services, taking jobs away from those US-born, leading to increasing violent crime, and diluting American values and identity; others contend that immigrants advance the US economy by creating a market for goods previously in low demand, increase new business entrepreneurship, and add to the work force by taking low wage jobs native workers do not desire. López and Bialik (2017) of the Pew Research Center estimate that of the 160.4 million in the labor force in 2014, 19.5 million were immigrants and 8 million were unauthorized immigrants. The same authors also make the distinction that lawful immigrants often perform professional jobs in management, business, or finance; unauthorized immigrants, however, are most likely to work in service or contracting jobs.

Some researchers hold that "immigrants inject vitality into the American economy; they bring their intellectual and scientific talents to United States research and development efforts, and help boost United States labor force productivity". Indeed, "a century of Latino migration, for example, has resulted in millions of law-abiding citizens who contribute to the U.S. economy and are as patriotic as any other American" (McGruder, 2016, p. 1). In fact, Minier (2017) noted that more than half of the 87 technology startups valued at over a billion were co-founded by immigrants. Minier (2017) further pointed to a letter signed by nearly 1500 economists, citing the numerous benefits that a constant flow of immigrants brings to the United States economy.

A National Academies of Sciences, Engineering, and Medicine (2017) study shows that "the mix of skills possessed by arriving immigrants – whether manual laborers, professionals, entrepreneurs, or refugees – will influence the magnitude and even the direction of wage and employment impacts". However, "when measured over a period of more than 10 years, the impact of immigration on the wage of natives overall is very small" (p. 4). Nonetheless,

Abramitzky and Boustan (2016) find that concerns of United States-born Americans continue to focus around fears that immigrants would not only lower their wages, but also that they may not integrate well into the social fabric. The concerns over wage and employment shortages are certainly not new. As Lee et al. (2017) describe, around the Great Depression – between 1929 and 1937 – almost half a million people of Mexican descent (many already American citizens) were forcibly deported to Mexico in order to save and create jobs for US-born citizens. Not only did it not work as hoped, it in fact backfired: in those places that sent away most Mexican-Americans, the job markets got smaller and unemployment rose.

As this discussion shows, both old and new views of immigration oscillate between benefits and detractions. So what should the role of immigrants be in the US, in the long run? This question has been asked for some time, and seems no closer to having a clear answer. It has, however, engendered two distinct schools of thought. The following section discusses the philosophical bases of these two approaches.

## Two Philosophical Approaches to Immigration

In its early days, the United States of America accepted immigrants freely. As a country of immigration from its origin, and given the low overall population, there were no immigration laws or restrictions to bar immigrants from entering the country. However, this changed in the early 1900s. The 20th century, without dispute, was the time when various ideas around immigration, its impacts on the American culture, and the acculturation processes of individuals and groups were most hotly debated. As a result, two competing schools of thought emerged. The viewpoint of the majority group demanded full Americanization and Anglo-conformity, which stipulated immigrants who chose the United States as their new home become part of the melting pot. This vision later became a "central element in the development of the assimilation school of race and ethnic studies in American sociology" (Hirschman, 1983, p. 398).

Advocates of the melting pot philosophy expected that as new immigrants arrived to the new land, they would become culturally and racially mixed to create an America 'Utopia'. Therefore, these 'conformists' set in motion to materialize this vision through the model of the melting pot. A model described by Gordon (1961) as Anglo-conformity demanded English language and English-oriented cultural patterns to be the standard way of life. It is important to note that the philosophy of the melting pot is mostly associated with the United States and its immigration history. Its process broadly associated cultural assimilation in which the melting of cultures and intermarriage of ethnicities were considered aspects of this philosophy.

The viewpoint of the melting pot led to the birth of the competing philosophy, which promoted acculturation over assimilation. The scholars who held this viewpoint advocated that immigrants maintain their original culture, ethical identities, customs and values; however, they would also adapt to the culture of the new environment. This school of thought saw the philosophy of the melting pot to be forceful and overly simplistic in its view towards the understanding of culture and ethnicity. This group of scholars advocated a multicultural society that is inclusive yet aware of cultural and ethnic differences. Further, they saw the majority's ideal of attaining a homogeneous America as a romantic vision that would strike most modern observers as naïve and rather patronizing (Hirschman, 1983). Multiculturalists advocated that immigrants and minority groups should be integrated into US society with little stress and little pressure. This set the ground for the debate that went on through the last century and that is not yet over.

## The Foundations of the Melting Pot Philosophy

According to Hirschman (1983), the term 'melting pot' was first used by Ralph W. Emerson in the 1800s. However, the term only came to general usage after it was used in a play with the same name describing a fusion of nationalities, cultures and ethnicities in the early 1900s. The melting pot, as a philosophy of the dominant group, advocated the view by which people of different cultures, races, religions, and ethnic backgrounds would come to form a homogeneous society that would become part of the larger multi-ethnic society. This was best exemplified by Robert Park, an influential figure from the University of Chicago, who developed the theory of assimilation concerning US immigrants while carrying out research work in urban Chicago communities (Hirschman, 1983).

In his theory of assimilation, Park (1950) introduced four steps to the race-relation cycle in the story of an immigrant: contact, competition, each group learning to accommodate the other, and, finally the immigrant group would learn how to assimilate into the dominant group's ways of life. This view became the theme of the melting pot: that immigrants would give up the identities that connected them with their country of origin, absorb the norms and culture of the United States, and would in turn be absorbed into the dominant culture. Hirschman (1983) pointed out Park's assumption that the forces of change in modem societies would ultimately obliterate divisions based upon irrelevant criteria of language, culture, and race; this vision would result in democratic political institutions, and the industrial organization of modern society would then recruit and promote individuals on the basis of merit and not ethnic origin. With this vision in mind, the dominant nativist group worked to set polices that would meet this model of the melting pot.

**The Migration of the Early 1900s**

According to McGruder (2016), countless immigrants came to the US in the 1600s mainly to escape religious persecution or find fortune in the new land; between the 17th and 19th centuries, however, hundreds of thousands of African slaves were brought to the US against their will. And starting around the mid-19th century there were significant Chinese immigration waves, which went from being tolerated to engendering the first legislation restricting a particular ethnic group from immigrating: the 1882 Chinese Exclusion Act.

In the early 1900s, the United States saw massive numbers of European immigrants entering the country. According to Suárez-Orozco (2000), 8,798,386 Southern and Eastern European immigrants, from Ireland and Italy for instance, arrived during the first decade of the 20th century. Even though these were not the kind of immigrants that Gordon (1961) identified as Anglo-conforming, or English and Protestant, the Act of 1790 allowed them to enter the nation freely.

The massive number of new immigrants may have been unusual ingredients for the melting pot to readily absorb. However, at that time, immigrants were seen both as laborers that filled gaps in the workforce, and as an economic boon that would lead to social growth. Gordon (1961) stated that, on one hand, these immigrants added to population numbers, worked on farms and in mines, and built the railroads; on the other hand, the poverty-stricken Irish Catholics and substantial influx of Germans became a source of anxiety to the overwhelmingly Protestant society.

Nonetheless, the melting pot proved large enough to accommodate these groups of immigrants. It may not have been right away, especially in the case of the Italians, but they were gradually accepted and became fellow 'white' citizens. However, to be fully integrated they had to learn the English language, United States work ethics, and other mannerisms. Schools like the Ford English School set out to teach these immigrants the English language in the name of making them better citizens, and protected them against the many pitfalls which lurked in the path of the unwary foreigners (Pozzetta, 1991).

Gordon (1961) mentioned that as far back as colonial times, Benjamin Franklin recorded concerns with German immigrants' slowness in learning the English language and especially with the establishment of their own native-language press. This type of behavior would of course contradict the dominant view of melting pot because Anglo-conformity demanded that these immigrant groups cast off their European skin, and gaze forward instead of dwell on their past (Gordon, 1961). Therefore, stripping them of their native language and cultural values were the first steps toward full memberships in the melting pot.

The philosophy of the melting pot also ran into trouble with trying to maintain the 'whiteness' of the dominant population. Italians were in fact not seen as white in the beginning, but eventually were absorbed into the existing culture. However, according to Gordon (1961), although Black Americans made up nearly one-fifth of the total population at the time, their predominant slave status, combined with racial and cultural prejudice, barred them from serious consideration as an 'assimilable' element of the society. Furthermore, measures were taken to fully guarantee that this undesirable population remain outside the melting pot.

Hirschman (1983) talks about how legal barriers, including the denial of the opportunity to vote among many other essential citizenship rights, kept Black Americans in a state of powerlessness, a practice described by Myrdal (1964) as a moral dilemma between American ideals of equality and the practice of racial discrimination. Where Native Americans were concerned, Gordon (1961) stated that assimilation was out of the question: they did not want it since they had a positive need for the comfort of their own communal institutions.

These questions about who is or is not able to – or is desirable to – be absorbed into the American melting pot have indeed never been resolved, and they are debated afresh with every new influx of immigrants. In fact, as Young notes: "In this nativist vision, the time period to which we return is one in which immigration is sharply restricted by national, ethnic, and religious criteria" (Young, 2017, p. 218).

### The Philosophical Perspectives of Multiculturalism

Scholars who felt that the simplistic philosophy of the melting pot did not address or consider the cultural and ethnic differences of all those involved advocated the concept of cultural pluralism. Some concerned scholars in the 1920s introduced this concept of the preservation of differences and pluralism of cultures, as opposed to the melting pot vision (Rasmussen, 2008). Multiculturalists recognized the inherent value of different cultures and the need to preserve them; they asserted that assimilation could hurt minority cultures by stripping away their distinctive features. For example, teaching the English language for the purpose of stripping immigrants of their native language and cultural values was seen by this group as inimical to the future of the United States (Hirschman 1983).

Gordon (1961) noted that World War I gave Anglo-conformity its fullest expression in the ideology of the Americanization movement: stripping immigrants of their native culture and attachments would make them Americans along Anglo-Saxon lines. The war was seen as the opportunity to speed up the dissembling and reassembling process into the melting pot.

However, the ideology of a multicultural society gained momentum in the 1940s. Indeed, the "nation of immigrants" concept was used in the 1950s to showcase the United States as a land filled with opportunities. Suárez-Orozco (2000) noted that the notion of conservative intellectuals seeking to safeguard Anglo-American cultural traditions was rejected by intellectuals on the left who embraced ideas of cultural pluralism. Indeed, multiculturalists claimed that the American mainstream was nothing more than an implicit oppressive mechanism that worked to subjugate those of minority groups.

Adepts of multiculturalism debated how best to approach the issues of immigration and national identity and laid the foundations for cultural pluralism. This, perhaps, may have come around full circle when Child (1943) and Lewin (1948) advocated acculturation as the strategic reaction of the minority in continuous contact with the dominant group. The psychologist Child (1943), when studying second-generation Italian Americans in New Haven, Connecticut, during the late 1930s, found that many Italian Americans lapsed into an apathetic identity state. The author found these populations to have loyalty to both identities. This view gave immigrants and minorities several options with different motivations and consequences in which they could choose to acculturate into the dominant group.

As a result, a wide range of issues involving race, economy, education, and the right to vote had to be given consideration. The views of the dominant groups were that these types of race/ethnic divisions would eventually disappear, or at least be minimized in industrial society, and democratic political processes would eventually override the forces of prejudice and discrimination (Hirschman, 1983).

We know the immigrants of the early 1900s were absorbed into the boom of the industrialized age; however, toward the end of the first half of the 1900s America was shifting from its industry-based economy to a competitive market system. Following World War II, therefore, there was a shift in the kind of workforce that the markets required and some softening in the melting pot requirements. The melting pot became, at a minimum, racially inclusive. But cultural pluralism was about to gain further momentum with the American Civil Right Movement and the enactment of the Immigration and Nationality Act of 1965.

The radical 1965 law abolished the old quota system based on national origins, and opened the way for new waves of immigrants of color, which had been barred from entering the country before then. According to Suárez-Orozco (2000), starting in 1965 a million new immigrants arrived in the United States every year until 1990; most of these were immigrants of color, the majority being from Latin America and Asia.

These were not the Irish or Italian white immigrants who arrived during the expansion of industrialization, and indeed Hirschman (1983) indicated that this change heightened the degree of racial antagonism overall. Hutchinson (1965) noted that from 1919 to 1950 there was a shift toward more skilled occupations among the foreign-born and their children. But it was not clear at the time how the arrival of these immigrants would affect the market, and therefore which philosophical frame – melting pot or multiculturalism – would prove a more useful lens, especially with regard to education and the workplace.

According to Suárez-Orozco (2000), it was no longer useful to assume that immigrants were joining a homogeneous society dominated by the middle-class, white, European American Protestants. The new immigrants of color were in fact reshaping the structures and foundation of the melting pot. It was this fundamental social shift that made schooling and education even more important for immigrants: education was a key indicator for socioeconomic achievement as well as an investment that influenced subsequent social and economic mobility (Hirschman, 1983). In addition, Suárez-Orozco (2000) notes that since the 1960s those coming to the US consisted of a mix of highly skilled and educated, along with low skilled immigrants. So, the melting pot assumption that held immigrants would do better in terms of schooling, health, and income only the longer they were in the United States was no longer valid.

Given this mix of backgrounds, immigrants tended to respond differently to the American way of life. On one hand, they did not need to be here a long time to do well, if they were highly skilled and highly educated. Others, however, may be poorly educated and unskilled, but still possess distinctive cultures and values that are complex and dynamic (Banks, 1999). Therefore, any forceful attempt at absorbing this latter immigrant group into dominant society may have led to what was known as acculturative stress. First noted by Redfield, Linton and Herskovits (1936), acculturative stress includes emotional reactions such as anxiety and depression. This was later theorized by Born (1970) and then Berry (1980) who found acculturative stress to be a significant problem for many immigrant groups.

## The Challenges of the Two Competing Philosophies

The theory of the melting pot seems to fit less well for how immigration has looked like since the 1960s. Most of these immigrants come from Latin America, Asia and Africa, and, unlike the Irish, Italians, and Germans, they do not spread out across the country. The US Census Bureau's 2000 publication indicated that the majority of these immigrants settle in only a few states. For example, seventy-one percent of all recently arrived immigrants to the US reside in the metropolitan areas of California, New York, Florida, Texas, New Jersey, and Illinois. Although this may be able, to some degree, lead to a 'melting-pot city', it by no means ensures a 'melting-pot country'.

Since new immigrant groups tend to settle in ways completely different than earlier ones, we have to ask if the new immigrants of color indeed are recreating the structures of Anglo-conformity. The reasons for these differences are not too difficult to imagine: crossing international boundaries leads to extensive life changes, including adapting to new cultural values, social rules, policies, and material environments; this can often be accompanied by a sense of loss in terms of careers, social ties, social status, and social identity (Tsai, 2006). This is perhaps more intensely the case for immigrants of color than it was for previous European immigrants, as the change in environment can be much greater. Therefore, to complement for these shortcomings and reduce their stress, these immigrants choose their settlement areas in ways that enable them to tap into existing support networks, leading to concentrations of various ethnic groups in only a few cities.

Suárez-Orozco (2000) noted that the US society is no longer, if it ever was, a uniform or coherent system. The immigrants of today, given their financial resources and social networks, end up gravitating toward very different sectors of the country. Given this reality, how can it be ensured that there is uniformity in the United State of America? It can well take several generations, if not longer, for these immigrants to completely blend into US culture. While the melting pot philosophy dominated, economic needs and economic pressure hurried ethnic integration; nonetheless, the process of alleviating pressure may have resulted into a society that is now more fragmented.

Prior to World War I, European immigrants were seen both as a strength and a source of prosperity, and the nation welcomed them with – mostly – open arms. This made it seem that the melting pot was real and accommodating; nonetheless, the concept of the melting pot has been criticized first for being slow to decide who was 'white enough' to be absorbed into the society, and later on for being unkind and racist to non-European immigrants. Indeed, perhaps the very foundational concept of the melting pot may be criticized directly, given that its vision of mixing various cultures as well as races was supposed to end up being, for the most part, a white Anglo Saxon culture.

It has been noted that since the 1960s most sociology and history research has increasingly overlooked the philosophical view of the melting pot in describing ethnic relations in the United Sates (Adams & Storther-Adams, 2001; Gordon, 1964; Glazer & Moynihan, 1970). Nonetheless, there have been some compromises between the two competing philosophical views discussed in this paper. For instance, this compromise is clearly seen in the use of English as the primary language in school, despite the importance of respecting students' first culture, and recognizing the equal importance of the adoptive (Goldberg, 1974).

Suárez-Orozco (2000) further noted that the progressive movement of multiculturalism is built on the Anglo-American culture and that its support for the equality of immigrant cultures made possible the further development of pluralistic ideas, to which the twentieth century espoused, as well as American cosmopolitanism. Yet there are still contradictions to be found in both schools of thought. The multicultural approach typically urges against the forceful assimilation of immigrants, and often supports bilingual education and affirmative action; however, it is still unclear whether the desired end results are a relatively homogenous society.

Also, the melting pot philosophy cannot be entirely dismissed either: as distinguished social scientists and others who advanced this field recognize, the common theme is that the immigrants of today can and will become Americans and that they will enrich this nation's life, like so many before them.

### The Scapegoating of Immigrants: Old Wine in a New Bottle

Using immigrant narratives to win elections, or blaming them for various local or national challenges is not a new phenomenon. Indeed, this is something that has been going on for over a century, and various groups of immigrants are by turns viewed with positive enthusiasm for their vibrant contributions, with distrust of their willingness to assimilate, or with downright rejection of their unusual backgrounds. Oftentimes, the legitimate challenges faced by many working and middle-class citizens have been weaponized against immigrants. Indeed, "socially disadvantaged groups, Betz suggests, are most prone to blame ethnic minorities and migrant populations for deteriorating conditions, loss of manufacturing jobs, and inadequate welfare services" (Inglehart & Norris, 2016 p. 11). Perhaps what is most notable in our current environment is the similarities with the nativist perspectives common around the turn of the 20th century, "particularly the focus on the purported inability of specific immigrant groups to assimilate, the misconception that they may therefore be dangerous to the native-born population, and fear that immigration threatens American workers" (Young, 2017, p. 217). Indeed, as Young (2017) notes, these ideas never really go away, but merely change from background to top of the agenda at various times, especially during economic downturns.

The immigrant demographic today is very similar to the US in 1920: 13.5 percent of all Americans today are foreign-born, a figure that stood at 13.2 percent in the early 20th century. And as in that earlier time, many doubt that the immigrants of the time could be assimilated into the main US culture. That is why now, as then, "politicians and the press frequently portrayed immigration as a threat to the nation" (Young, 2017, p. 218). The difference of course is that around the turn of the 20th century most immigrant groups were white Europeans. Currently however, most immigrants are Latinos, Asians and Africans. They are both documented and undocumented immigrants, and often their combined circumstances make it more difficult for them to assimilate.

Also notable at this current time is the reminder that many in the United States perhaps assumed that, regardless of previous immigration, the country would remain mostly white and Anglo-Saxon. Back in the early 1950s, Senator Patrick Anthony McCarran, a Democratic Senator from Nevada – who co-sponsored the 1952 Immigration and Nationality Act bill – argued that assimilation (i.e., the ideal outcome) would only take place if the number of immigrants were kept in check and controlled by established 'old-stock' Protestant and Catholic Americans (Gerstle, 2017). Gerstle (2017) notes that the viewpoint of many white ethnics continues to be that all newcomers "assimilate to Anglo-Saxon ideals" (p. 330).

Therefore, what we saw before and after the 2016 US election was not new, it was simply using immigrants as a scapegoat in the face of the domestic economic downturn. "In the past, much like today, politicians accused immigrants of maintaining distinct cultural norms, continuing to speak foreign languages and living in enclave communities" (Abramitzky & Boustan, 2016, p. 1). During the 2016 election, nominee Donald Trump focused heavily on immigration restrictions including building a wall along the US Mexico border, and ensuring that Muslims from some countries cannot enter the US (Rothwell & Diego-Rosell, 2016). And it was a lot of that type of populist rhetoric that was "especially appealing to lower middle class White people who felt disenfranchised and displaced" (Berlet & Lyons, 2018, p. 17).

Indeed, the economic hardship felt throughout the country does affect some groups more than others, and among those hardest hit are workers in low-wage jobs, those in cyclical industries, and poor whites living in areas with high immigrant concentrations. In addition, the current rhetoric does not only focus on anti-immigration policies, but also brings in traditionally left-leaning ideas of protecting various social benefits, such as social security, or economic priorities, such as infrastructure. Taken together, it is not hard to see why disadvantaged groups respond so well to harsh rhetoric that scapegoats immigrants and puts the blame of lost job opportunities and public services on them (Inglehart & Norris, 2016). However, in another similarity to anti-immigrant times in our history, current talking points use fear and misinformation to turn feelings of dissatisfaction and marginalization into political drive (Daftary, 2018).

As previous sections of this paper have shown, immigrants both documented and undocumented help sustain our economy. Reston (2015) reported that immigrants give more to the US welfare system than they receive. For example, on earnings of $240 billion, they have payed $90 billion in taxes and used only $5 billion in public benefits. Likewise, in its 2013 report, the US Chamber of Commerce found that although undocumented immigrants do not tap into the US welfare system, they do pay federal and state income, Social Security, and Medicare taxes. Further, Warren and Kerwin (2017) noted that almost a million undocumented immigrants own homes and pay mortgages.

Even so, many of those who connected with President Trump's immigration rhetoric have legitimate grievances. No country in this world will survive with uncontrolled borders. The economy has steadily become more knowledge-based and global in nature. Well-paying manufacturing jobs that once made it possible for so many to live the American dream can now be done either by automated systems, or by people living around the globe and being paid fractions of the salaries companies would have to pay Americans. It is also true that the United States does tolerate illegal immigration. No other country on earth has 11 million undocumented immigrants. However, this is not by accident. The US economy has learned to rely on both documented and undocumented immigrants to thrive. Undocumented immigrants, in particular, gladly preform many jobs that most Americans otherwise would not. Among others, they build our homes more cheaply and keep our grocery bills low, when compared to other developed nations.

Finally, as shown earlier, immigrants do affect wages and employment in the short run; so there is certainly validity to arguments about some job opportunities not being available to US-born Americans. However, at the end of the day, it comes down to supply and demand. Therefore, finding long-term solutions to these legitimate economic problems would require genuinely addressing the issue of supply and demand and looking at how the economy has been transformed; otherwise, immigrants will continue to be blamed for things far beyond their control.

## Conclusion

From the above discussion, it would seem that advocates for the melting pot philosophy would like to see the United States as a nation of immigrants with strong nationalism, one people, one culture, and one language – English. However, it is also fair to say that the melting pot approach has shifted from previous goals of stripping immigrants of their identity on arrival. This may partly be due to occupational and workforce demands. There are some jobs that need to be filled, and which will ordinarily not be filled by 'people within the existing melting pot'. Low skilled immigrants are often willing to take up those jobs.

The demographer Myers (2007) notes a couple of powerful demographic shifts at work. The large inflows of immigrants during the past three decades are seen with enormous levels of anxiety. This stems unsurprisingly from the fear of new and unwanted changes, especially to the nation's ethnic, social, and economic identities. At the same time, the retiring baby boom generation is draining the nation of highly skilled workers. Myers (2007) believes that the former shift can help solve the latter, as immigrants have the potential to fill the gaps forming in the workforce.

The question remains, however, whether the melting pot ideal still works for the changing US landscape. Where the US culture was once the force changing those who came to call it home, recent demographic changes and new waves of urban-bound immigrants are the ones who are in fact transforming the nation. Philosophers and sociologists may argue that the concepts of the melting pot mean little more than Anglo-conformity, which means that smooth assimilation would be nearly impossible for a majority of immigrants. Indeed, the study of culture is a complex undertaking. It gets even more complex if there exist a widely held assumption that 'cultural capital' is the culture of the dominant group in a society. Promoting the integration of immigrants into host communities may be more challenging when, in fact, there are no immigration policies and federal laws in place that explicitly support promotion of social, economic and civic integration in the US, despite the fact that the country is shaped by immigration (Gozdiak & Martin, 2005).

So does that mean the multiculturalism is the future of the US? On the one hand, Payan (2016) says that "many see the increased diversity that comes with immigration as a threat to national identity" (p. 1). On the other hand there is a definite cultural transformation happening in the US: younger and college-educated groups in particular are showing increased tolerance towards all sorts of diversity, including LGBTQ+ rights, diverse family groupings, gender identity, diverse habits and ethical norms, and multicultural lifestyles (Inglehart & Norris, 2016).

Hirschman (1983) notes that education is an achievement in the socio-economic hierarchy as well as a resource that influences subsequent social and economic mobility. As such, immigrants' education is seen as the primary step toward full participation in American society. This may lead to immigrants becoming financially secure and fully participating in United States culture. However, this cannot be left to the democratic machine or market to figure out. Let us remember that anti-immigrant fears at the turn of the 20th century drove forward legislation that sharply decreased the number of immigrants to the US for many decades to come. The current policy discussion seems to be lost in translation of these two philosophies. To be able to develop a policy that will be beneficial and lasting, a good understanding of these two contrasting philosophies – and their merits and demerits – is crucial.

We can all likely agree that if the desire is to stem the flow of migration to the US, then scapegoating immigrants and using the legitimate hardships faced by working- and middle-class Americans against their follow citizen will not bring about a solution to the legitimate immigration issues in this country. Therefore, we believe that policy should focus on the following.

First, any immigration policy should start with the fact that the United States today is a multicultural nation. Moreover, as long as we have an interconnected global economy, and given the fact that people will always migrate toward a better life, the United States will remain a multicultural nation.

Second, for immigrants to fully contribute to the US society and economy, immigration policy needs an emphasis on education—investment in education that goes beyond basic English language literacy skills that will lead to entry-level jobs. By all measures, education remains a key indicator for socioeconomic achievement as well as an investment that influences subsequent social and economic mobility. Therefore, given the fact that today's economy is global and highly techno logical, we believe any future immigration policies should have a greater emphasis on education.

Third, domestic immigration policy has to recognize global economic, labor, and trade policies. Overall, immigration has often shown "positive effect on trade flows," and as such, trade arrangements "to create legal migration opportunities" can curb the flow of migration in general and would allow the kind of immigrants needed for specific sectors of the economy (Schmieg, 2019). Moreover, according to Budiman and Connor (2019) of the Pew Research and based on World Bank data, globally immigrants sent $625 billion to their home countries in 2017. Likewise, immigrants send more than $148 billion to their perspective countries from the United States. Therefore, appropriate policies that support this type of targeted migration could have a great impact to local (source) economies, hence playing a role in curbing migration.

In fact, there are empirical findings going back a century and highlighting various positive and negative implications of migration – whether on employment or the economy as a whole. To sustain today's economy and meet tomorrow's challenges – such as our ageing population – immigrants are critical to all sectors of our economy. It is likely that the flow of immigrants will remain both highly educated, mobile, and globally-minded; they will likely keep bringing with them both intellectual and monetary capital to this county, and help the US maintain their competitive edge on the global arena. Likewise, immigrants who come to this country as a result of political oppression and economic disparities are mostly low skilled individuals who are happy to perform tasks that Americans overlook and the labor market demands. In totality, immigration is neither good nor bad. What is good or bad is the right or the wrong, or missing, immigration policies.

There is the need to take an honest look at the political context US immigration is situated and develop plausible strategies for dealing with it. Otherwise, issues will always be lost in translation. As Triadafilopoulos (2010, p.129) aptly puts it, any immigration policy, "no matter how well conceived and reasonable, will be subject to intense debate and modification, as a consequence of interest group lobbying, political entrepreneurship and the normal dynamics of American politics."